%% file: sample-authordraft.tex
\documentclass[sigconf]{acmart}
\usepackage{graphicx}
\usepackage{booktabs}
\usepackage{url}
\usepackage{hyperref}
\usepackage[hyphenbreaks]{breakurl}
\AtBeginDocument{%
  \providecommand\BibTeX{{%
    \normalfont B\kern-0.5em{\scshape i\kern-0.25em b}\kern-0.8em\TeX}}}

\copyrightyear{2024}
\acmYear{2024}
\setcopyright{rightsretained}
\acmConference[PDC '24 Vol. 2]{Participatory Design Conference 2024, Vol. 2: Exploratory Papers and Workshops}{August 11--16, 2024}{Sibu, Malaysia}
\acmBooktitle{Participatory Design Conference 2024, Vol. 2: Exploratory Papers and Workshops (PDC '24 Vol. 2), August 11--16, 2024, Sibu, Malaysia}\acmDOI{10.1145/3661455.3669867}
\acmISBN{979-8-4007-0654-7/24/08}

\begin{document}

\title{Reconfiguring Participatory Design to Resist AI Realism}

\author{Aakash Gautam}
\affiliation{%
  \institution{University of Pittsburgh}
  \city{Pittsburgh}
  \state{PA}
  \country{USA}}
\email{aakash@pitt.edu}








\input{Paper_Content/0_abstract}



\maketitle

\input{Paper_Content/1_introduction}
\input{Paper_Content/2_literature_review}

\input{Paper_Content/3_reals}
\input{Paper_Content/5_discussion}
\input{Paper_Content/6_conclusion}

\begin{acks}
 I am grateful to Aditya Nayak and Deborah Tatar for their feedback. 
\end{acks}

\bibliographystyle{ACM-Reference-Format}
\bibliography{sample-base}

\end{document}

%% file: Paper_Content/0_abstract.tex
\begin{abstract}    

The growing trend of artificial intelligence (AI) as a solution to social and technical problems reinforces AI Realism---the belief that AI is an inevitable and natural order. In response, this paper argues that participatory design (PD), with its focus on democratic values and processes, can play a role in questioning and resisting AI Realism. I examine three concerning aspects of AI Realism: the facade of democratization that lacks true empowerment, demands for human adaptability in contrast to AI systems' inflexibility, and the obfuscation of essential human labor enabling the AI system. I propose resisting AI Realism by reconfiguring PD to continue engaging with value-centered visions, increasing its exploration of non-AI alternatives, and making the essential human labor underpinning AI systems visible. I position PD as a means to generate friction against AI Realism and open space for alternative futures centered on human needs and values.

\end{abstract}

%% file: Paper_Content/1_introduction.tex
\section{Introduction}
\label{sec:intro}




``It's easier to imagine the end of the world than the end of capitalism'', writes Mark Fisher \cite[pp. 1]{fisher2022capitalist} to illuminate the change in global polity, which since the fall of the Soviet Union around 1991 has portrayed that there is no alternative to capitalism. Fisher calls this phenomenon ``Capitalist Realism'' \cite{fisher2022capitalist}.
In the era of capitalist realism, every aspect of our lives is governed by capitalist rules and values.
At the same time, various mechanisms for shaping a collective imagination that does not see a possibility beyond capitalistic approaches are deployed, including conflating capitalism with democracy, emphasizing the power of political participation, and contrasting the current order with tyrannical socialist governments that failed. 

We have also entered the era of AI Realism, where problems that were previously attempted to be solved collectively through social and technical mechanisms are now proposed to be solved through corporate-controlled AI (c.f., \cite{ castelvecchi2022chatgpt,clayton2024ai}).
The ``AI'' I talk about here is not the statistical approach but rather the much-hyped larger ecological system that is being pushed by corporations.
The growing use of generative AI systems in different workplaces \cite{castelvecchi2022chatgpt}, academic venues' rules on the use of ChatGPT or other large language models \cite{acmPolicy}, and the growing use of chatbots by existing services \cite{aljazeeraAiTherapy} all point to the fact that we have arrived at AI Realism.
The shift to AI realism has been more rapid than that of capitalist realism though. 
The success of AlexNet in 2012 began a drive to push AI as \emph{the} solution to a range of problems \cite{christian2020alignment}, and by 2022, with the release of ChatGPT, it had captured the public's imagination as a panacea (e.g., \cite{de2023chatgpt, aljazeeraAiTherapy, extance2023chatgpt}).
In some cases, there have been concerted efforts to push AI as a solution, in others, we inadvertently accepted it. 

More than the striking parallels, Capitalist and AI Realism are closely connected. 
AI is a tool that existed long before 2012, but \emph{the} AI that is being pushed widely is more than a statistical tool. 
It is the tool as used by and defined by the capitalist; a technological mechanism that enables the concentration of resources and power, hidden behind the veil of neutral-seeming statistical formalism\footnote{Although statistical-AI and AI-as-capitalist-instrument are tied to some extent considering the corporate control over the data and other infrastructure \cite{whittaker2021steep}.}. 
The pervasive promotion of AI with claims of efficiency, productivity, and various other capital value creation --- referred to as ``AI snake oil'' \cite{narayanan2019recognize} --- has distracted us from examining details of the value and the potential for harms that the system brings.

In the face of growing AI Realism, PD has a significant role to play. 
To illustrate this, I begin by questioning three aspects of AI that are promoted widely. 
First, the claims of ``democratizing AI'' lack the necessary distribution of power to the people affected by the AI system to be democratic. 
Second, the popular imagery of AI for personalization makes significant demands of human malleability and erodes opportunities for collective action. 
Third, the image of AI systems as efficiency-promoting obfuscates the labor required to enable them.
Reflecting on these three aspects, I argue for reconfiguring PD methods to resist AI Realism. 
These reconfigurations include contesting narrow metrics to enable broader visions of the future, incorporating explorations of non-AI alternatives while identifying possibilities, and making visible all the human labor that enables AI systems. 
In doing so, I seek to (re)position PD as a force for setting agendas and upholding democratic participation, and, with it, centering human values in the era of AI Realism. 
\\


%% file: Paper_Content/2_literature_review.tex
\section{Literature Review: AI and the Core Tenets of PD}
\label{sec:lit}
Participatory design (PD) finds its roots in the 1970s with the workers' push for greater control and democratization amidst the introduction of new technology in the workplace \cite{bodker2018participatory}.
Over the years, PD scholarship, within the workplace and beyond it, has highlighted tenets like carefully developed participation, mutual learning between workers and designers, and envisioning alternative technological futures \cite{ehn2008participation}. 
These principles support creative cooperation in designing new sociotechnical systems, thus enabling potential space to challenge existing power dynamics \cite{bratteteig2012disentangling, muller2012participatory}.
In some respect, the wheels of time have brought us back to the 1970s with the advent of AI in the workplace and beyond, with growing concerns of drastic changes in work practices and values \cite{brookings2019}. 

AI systems bring a set of unique characteristics, some common with other large-scale technologies and a few unique to it. 
Among the unique ones, AI systems often lack transparency in how it was created and evaluated, in data collection, model development, fine-tuning, and impact evaluation \cite{pushkarna2022data, gebru2021datasheets, raji2020closing, sambasivan2021everyone}.
Second, AI system development is expensive and typically deployed on large scales, thus requiring infrastructure that only a few can afford \cite{whittaker2021steep}. 
Third, AI as a ``modern technology'' is justified for a fast-paced push, where rapid development and deployment evades auditing, evaluation, and even regulations. 
In this sense, the proliferation of AI embodies Silicon Valley's ethos of ``move fast and break things'' \cite{taplin2017move} or ``ask for forgiveness, not permission'' \cite{davies2019ask}.
Given its pervasiveness, we are all being participated in its growth by providing it with data; at times, there seems like there is no alternative. 

Efforts have been made to promote transparency and create possibilities for broader group engagement (e.g., \cite{ pushkarna2022data, gebru2021datasheets, mitchell2019model, raji2020closing}). 
But these are technological solutions that assume AI is inevitable; the challenge of AI is a ``social, cultural, political, and ethical one'' \cite[pp. 2]{loi2018pd} where the technical decisions intermingle with social responsibilities.
The majority of participatory AI movements have been devoid of \emph{empowered} engagement of broader stakeholders, including the possibilities to challenge existing power differences or opportunities for recourse and refusal \cite{ delgado2023participatory, birhane2022power, bratteteig2018does, williams2022exploited}. 
They fail to illuminate broader and alternative futures, thus further solidifying the position of existing AI as the inevitable ``natural order''. 
In such cases, Fisher contends, we ``must reveal what is presented as necessary and inevitable to be a mere contingency just as it must make what was previously deemed to be impossible seem attainable'' \cite[pp. 17]{fisher2022capitalist}. 
Revealing these alternative possibilities and enacting moves towards realizing them ---i.e., resisting AI Realism---is where PD can play a significant role. 
\\


%% file: Paper_Content/3_reals.tex
\section{The Real Behind the Promoted Reality of AI Realism}
\label{sec:reals}


\begin{table}
    \caption{Contrast between the promoted reality and the Real in AI Realism}
    \begin{tabular}{|l|l|} \hline 
        \multicolumn{1}{|c|}{\textbf{Promoted reality}} & \multicolumn{1}{c|}{\textbf{Real}} \\ \hline
        \begin{tabular}[c]{@{}l@{}}Democratized AI with \\ transparency\end{tabular} & \begin{tabular}[c]{@{}l@{}}Lack of data ownership \\  and control\end{tabular} \\ \hline
        \begin{tabular}[c]{@{}l@{}}Personalized \\ recommendations\end{tabular} & \begin{tabular}[c]{@{}l@{}}Human malleability, \\  machine intransigence\end{tabular} \\ \hline
        Automation and Efficiency & \begin{tabular}[c]{@{}l@{}}Human labor behind the \\  infrastructure\end{tabular} \\ \hline
    \end{tabular}
    \label{tab:real}
\end{table}

Fisher, drawing from Lacan \cite{lacan1988seminar}, argues for invoking the `Real', that which threatens the promoted ``reality''. The Real is the ``traumatic void that can only be glimpsed in the fractures and inconsistencies in the field of apparent reality'' \cite[pp. 18]{fisher2022capitalist}. 
For Fisher the Reals that can be presented to challenge the realities in capitalist realism are ongoing and increasing environmental catastrophes challenging the growth fetish of capitalism, the burgeoning mental health crisis that undermines the push for privatization of stress, and the proliferation of bureaucracy in capitalism despite the widespread claims of arriving at market-driven efficiency \cite{fisher2022capitalist}.  
Drawing parallels, I share three Reals that could open avenues to question the reality promoted in AI Realism (Table \ref{tab:real}).

\subsection{Participation Without Power Is Not Democratic}

In many Participatory AI efforts, participation focuses on inclusion for design input rather than a critical mechanism to support people in setting an agenda. 
For instance, on reviewing 80 participatory AI projects, \citet{delgado2023participatory} found that only three projects involved stakeholders throughout the design process, and only ten involved them in setting the scope and purpose of AI. 
The effect of this is that those efforts presuppose AI as a given with its claims of benefits unchecked; participation is constrained to small-scale inputs on improving the development and deployment of AI. 

From this perspective, participatory AI is deployed as a mechanism to co-opt human labor toward acquiescence \cite{birhane2022power, ahmed2020we}. 
This weakened participation stems from perverse incentive structures benefiting corporate AI --- they benefit from expanding AI adoption with the facade of people's participation.  
There have been exception cases of community ownership and governance of data practices and AI systems, including deciding whether to use AI (e.g., \cite{cofey2021maori}).
However, constraints such as limited time and resources are often cited as barriers, leaving open questions of \emph{how} communities can be empowered in governing AI systems. 
Indeed, the opacity and rapid evolution of AI systems raise questions on what even constitutes meaningful participation. 

\citet{bratteteig2018does}, discussing if AI makes PD obsolete, argue for participation in (1) envisioning ideas with knowledge of what AI can and cannot do, (2) concretizing design ideas by engaging with values and visions, and (3) evaluating the impact of the decision. 
While these are important steps, they are not sufficiently political. 
Knowing the capabilities, which is a motivation for explainable AI, does not challenge the techno-deterministic stance that it promotes; participation is co-opted to attend to technical solutions conveniently solved through AI rather than engaging in institutional or social possibilities. 
Similarly, discussing higher-level values and visions only has the risk of arriving at consensus on systems that in reality may necessitate surveillance and other large-scale data collection mechanisms that the participants may not be aware of. 
Transparency and co-designing processes involved in developing and deploying AI systems needs to be added to the process of making collective decisions about AI. 
Further, for participants to evaluate the impact of AI systems especially in the design phase is challenging since effects are unknown until development and, often, seen after long-time deployment \cite{bratteteig2018does}. 
Thus, participation should involve collectively establishing and bringing into practice accountability structures before the development or deployment of AI technologies.


\subsection{Personalization Demands Human Malleability and Erodes Mutual Learning}

Personalization is one of the most visible aspects of AI. 
Marketing rhetoric pushes personalization as a universally desired feature, with one McKinsey report declaring, ``Consumers don’t just want personalization, they demand it'' \cite{mckinseyHyperPersonalization}. 
Despite widespread concerns about privacy and lack of awareness regarding the types of data being collected \cite{liao2021should}, AI systems continue to collect data for personalization, which indicates users' limited agency in challenging this juggernautal drive towards personalization.

Personalization surfaces two tensions with PD.   
First, personalization in AI systems demands human malleability.
This is not a new phenomenon: people accommodate technology by changing their behavior and practices around the technology \cite{suchman2007human, liao2018all}.
But with AI, we are required to accommodate it both while using it and, since AI is trained in use, in training it.
\citet{gyldenkaerne2020pd}, for example, in studying AI use in electronic health records in Denmark, highlight how an AI system in providing personalized insights imposed a new process of data collection even though the clinician's existing processes were sufficient for their purposes. It not only changed the way clinicians recorded patient data but since it was rife with ambiguity, it made additional demands on the clinicians to attend to the data processes, moving them away from the primary purpose of their work. 
Moreover, human flexibility demanded in training the AI reifies AI's assumptions about us and our practices, leaving little room for us to change those. 
Going back to the case of the clinicians, despite the mismatch in their practices and the new AI-based processes, they did not have room for recourse. Instead, they had to devise additional workarounds \cite{gyldenkaerne2020pd}.
This challenge of ensuring users' control over their tools and their environment is a classic problem that PD has grappled with \cite{bratteteig2016participatory}, but now with a much-pervasive technological system.

Second, personalization creates its own context, influencing the user to think about aligning to the demands of the system rather than finding opportunities to make common ground with others.
It raises challenges for surfacing shared experiences and engaging in collective action.
We are forced to work on pushing back on the limitations and impositions of our own spaces rather than drawing commonalities among a larger mass to push against the system.
This echoes the Marxist notion of ideology --- that AI systems are \emph{the} thing --- and acquiescing the alienation that it brings.  
A significant exercise of abstraction is necessary to arrive at shared experiences and opportunities for mutual learning; these skills are not easily accessible to all.
Personalization seeks to make participation individualistic.
For PD, which relies on forming publics --- of people forming attachments to shared matters of concern \cite{dantec2013infrastructuring} --- this poses a challenge.

\subsection{Claims of Efficient AI Hide Away the Hidden Labor}

Proclamations of AI's potential to drive efficiency and unprecedented scale abound.
Hidden behind the veneer of magical-seeming machine intelligence lies the indispensable critical infrastructure of human labor engaged in essential tasks of collecting, cleaning, and labeling data. 
This is not accounted for in the claims of AI's efficiency! 
For instance, the ImageNet dataset which was the foundation for the AlexNet model in 2012, required upwards of 25,000 people (Amazon Mechanical Turks) to build the dataset, with more than 45\% of them making less than USD\$5 a week \cite{li2010crowdsourcing}.
This labor is rarely counted.

The labor is typically sourced out to the Global South where worker conditions and compensations are, at best, questionable (e.g., \cite{wang2022whose, williams2022exploited, gray2019ghost}).
The indispensable human labor is obfuscated behind claims of technological progress of AI.
Paradoxically, the very AI that relies on the labor of many is employed to justify efficiency drives, leading to layoffs elsewhere \cite{thorbecke2024tech}.
A prominent example of this is the recent Duolingo layoff, where workers at most precarious positions were laid off citing AI-enabled efficiency\footnote{\url{https://www.washingtonpost.com/technology/2024/01/10/duolingo-ai-layoffs/}}.

There are two facets to the problem.
First, the promotion of values around efficiency and scale brought by AI normalizes the practice of hiding and often exploiting large human labor \cite{gray2019ghost}. 
The growing reliance on labor from the Global South enabled through complex sub-contracting practices has created mechanisms to evade accountability.
Those who use and are affected by AI are placed at a distance from those who build the infrastructure for that AI, enabling very little space for collective action and solidarity \cite{williams2022exploited}.  

Second, ``efficiency'' is presented as the only metric that should guide our endeavors. 
It has broader societal implications.
Examining the reality of what kind of labor is hidden and what kinds of work continue to be prized, we notice that the push for efficiency is top-down. 
The acceptance of efficiency requires acquiescing to unilateral decisions.
It leaves no room to explore alternative visions (e.g., promoting adaptability or interconnectedness) that we could collectively design \cite{huybrechts2020visions}, even when efficiency may conflict with our collective well-being \cite{schwartz2020why}. 
These two facets --- hiding the labor and leveraging the labored artifact to squeeze greater efficiency --- enable those with the power to create and deploy AI tools to concentrate wealth and resources.


%% file: Paper_Content/5_discussion.tex
\section{Discussion: PD in the Era of AI Realism}
\label{sec:discussion}
PD, with its emphasis on forming attachments to matters of concern \cite{dantec2013infrastructuring}, is uniquely positioned to question the relentless push for AI and create space to determine \emph{if and how} AI should be developed and deployed.

AI is not a modular unit; it is ecological and does not function in isolation. 
It is not monolithic either. 
Each AI system taps into and brings with it a myriad of standards, processes, and values, each of which in turn influences our actions and values. 
However, AI Realism pushes an instrumental and utopic view of AI, focused solely on outcomes (e.g., efficiency and growth) while obfuscating the complex processes and tradeoffs involved. 
In contrast, PD engages with processes that emphasize forming publics for collective envisionment, surfacing diverse values, and negotiating decisions over the technologies that impact them. 
More generally, there is an inherent tension: PD acknowledges diverse human conditions and seeks to support pluralism and individual agency in collectively realizing a shared vision of the future, whereas AI Realism seeks to --- or rather requires --- imposing AI as \emph{the} solution, removing it from the nuanced contextual differences and uncertainties inherent in our human condition.
This oppositional stance of PD can be constructive, allowing it to steer the relentless drive of AI Realism toward values that matter. 

Research on agonistic pluralism in PD (e.g., \cite{bjorgvinsson2012agonistic, disalvo2015adversarial}) provide guidance on creating the necessary friction to realize positive change. 
Scholars argue for embracing an adversarial stance rather than avoiding conflict. It is needed in the face of AI Realism.
Drawing from it,  I propose three initial reconfiguration of participatory methods to create space for resisting AI Realism.   
First, contesting dominant metrics to support engagements with broader visions and possibilities for the future. 
Second, including possibilities of developing non-AI solutions as alternatives to distill needs and situations that truly require AI solutions. 
Third, surfacing and engaging with the necessary human labor that is required to make the AI system function before making decisions on incorporating AI. 
Embracing these would allow participatory approaches to move beyond questioning whether AI works or how to make it work, towards surfacing systemic issues, assumptions, and impositions that underlie AI Realism.

\subsection{Beyond Dominant Metrics: Engaging in Value-Centered Visions}
The incessant focus on scale and efficiency --- a root element of modern-day capitalism --- finds fuel in AI systems. 
These metrics fail to capture the complex values that we care about.
Yet, in the rapid pace of incorporation, we rarely have an opportunity to question what values AI systems enforce and whether we want them.

There is a need to design alternative visions collectively. 
I echo \citet[pp. 7]{huybrechts2020visions}, who posit that ``designing ‘visions’ can turn the tension between addressing the focus on the big issues and the close attention to the particularity of relations into a dynamic dialogue that can repoliticise design.'' 
However, given the dominant focus on scale and efficiency, there is a need to first contest those metrics and discuss alternatives such as commoning and post-growth models (e.g., \cite{sharma2023post, teli2020tales}).

As we engage in designing for alternative visions, it is equally important, given the global pervasiveness of AI systems, to engage both with the local priorities and the global realities \cite{bodker2018participatory}, acknowledging and attending to the potential impacts, tradeoffs, and tensions at play across these levels \cite{gray2019ghost, crawford2021atlas}. 
Moreover, AI is not a monolithic system. Deconstructing AI into its constitutive parts is necessary to understand the vision it embodies. 
To this end, we may, for example, bring together the precarious workers early in the AI development pipeline and the higher-paid, slightly empowered workers --- the recent layoffs suggest no worker is truly in control --- in the design process of alternative visions and possibilities.

\subsection{Plural Possibilities: Exploring Non-AI Alternatives}
Moving past the dominant metrics is an important first step. 
A subsequent step along the line involves ensuring that PD methods make space for participants to enact their visions through non-AI alternatives. 

Specifically engaging with non-AI alternatives is important for the very act of co-designing AI systems, even when they are critical of other existing AI systems, enforces the belief that AI is necessary or inevitable.
Exploring non-AI alternatives challenges the assumptions that AI systems are better or necessary.
Further, the perception of AI as a widely-encompassing solution has made it easier to brush aside organizational and/or social problems. 
Engaging in non-AI alternative solutions may surface issues that cannot be fixed by technology. 

While exploring non-AI alternatives may not necessarily lead to the refusal of AI, it will afford space to critically reflect and distill aspects of the participants' needs and identify situations where AI is truly needed. 
In doing so, participants can deepen their understanding of the situation, and if AI is deployed, critically understand the role the AI system plays in the solution. This can help them gain some power over the AI systems. 

Moreover, exploring non-AI alternatives opens possibilities to reflect on what would be lost when a situation is reformulated to make it amenable to AI ecosystem. 
Scholars warn about ``algorithmic governmentality,'' where only outcomes that can be measured and realized through algorithms will be pursued \cite{nayak2021boredom, rouvroy2013algorithmic}. 
It is a reductive move, taking us away from uncertainties and politics to perfectly objective outcomes with no room for deliberation and recourse \cite{rouvroy2013algorithmic}. 
Exploring non-AI alternatives enables space to engage with uncertainties and the inherent messiness in our collective visions and values, potentially leading to plural possibilities of being.  

\subsection{Delegation Inversion: Making Visible the Human Labor}
Earlier, I presented two stages where human labor was essential to make the AI systems function: in creating AI systems such as with data labeling work, and our efforts to accommodate AI despite its rigidity. 
Both these essential labor are hidden from the popular narrative of AI. 
PD methods should surface and engage with it, particularly making it central in the decision-making process when participants are exploring AI-based possibilities. This echoes scholars who have long argued for making work visible \cite{suchman1995making, gray2019ghost}.  

For PD to resist AI Realism, we have to surface and engage with the realities of the hidden labor at the time of design envisionment. 
To enable this, I build on Latour, who, using a door as an exemplary technology, expounds on the value of ``delegation'', i.e., the work the technology does that humans no longer have to do \cite{johnson1988mixing}. 
I contend that for pervasive systems like AI, PD methods can be configured to engage participants in ``delegation inversion'' --- cataloging the additional work and sacrifice humans have to make for the technology to function. 
For instance, participants could create a ``labor accountability card'' to record information such as the number of data labelers, total labeling hours, and the working conditions of those contributing to the system supporting their envisioned AI solution.
Similarly, to enhance shared experiences regarding human adaptability, participants could note places of friction in using existing AI systems and where they needed to be flexible. 

These efforts of surfacing and engaging with the labor will help illuminate the cost of developing and deploying AI solutions, thus seeding informed decision-making on whether it is worth it. It can potentially create space for enacting alternative visions that do not demand such cost. 
\\

These three reconfigurations of PD methods --- contesting narrow metrics while envisioning possibilities, incorporating explorations of non-AI alternatives, and illuminating the often-hidden labor enabling AI --- aim to position PD as a powerful mechanism to resist AI Realism. They surface societal costs incurred by AI Realism, while simultaneously empowering people to shape or refuge technologies based on the values they care about.

However, this is a conversation at an early stage of PD's evolution in the era of AI Realism. 
There is critical work ahead for the PD community, both in research and practice.
We, as PD researchers and practitioners, must continue to challenge the dominant drive of AI solutionist approaches and ensure that the public is empowered to enact the future that \emph{they} desire.  
It is, indeed, a call for us to push back on AI Realism toward democratic, caring, and just futures shaped through collective decision-making. 
The stakes could not be higher. 



%% file: Paper_Content/6_conclusion.tex
\section{Conclusion}
\label{sec:conclusion}
The fundamental question of AI Realism is about making decisions on the distribution of resources and power. 
It is asking, ``Who should have the decision-making power on whether, what, how, when, where, and why AI systems should shape our collective future?''
As it stands, the power differences are highly skewed towards those who already control AI, and challenging it --- i.e., resisting AI Realism --- is where I see PD play a significant role. 
PD has always been political, but in the face of AI Realism, we need a vigorous renewal of the political. 
I argue for repositioning PD as a mechanism to create friction in the incessant focus on outcome that AI Realism pushes by engaging with the processes enabling AI and the values inherited by adopting it. 
Resisting the relentless push for AI adoption is an attempt to prioritize human values and justice over the corporatization and centralization of resources and power.  
As such, this work attends to the theme of PDC 2024 in arguing to extend the notion of participation beyond the specific methods towards being a driver for setting agendas in the face of AI Realism and establishing initiatives to reflect and facilitate alternative futures.
This calls for an alliance of researchers and practitioners who see PD as a democratic force to look outward and leverage PD methods to address root issues in the era of AI Realism, rather than inward at refining our methods.
